Metallic and nonmetallic double perovskites: A case study of $A_2FeReO_6$ (A= Ca, Sr, Ba)

J. Gopalakrishnan*, A. Chattopadhyay, S. B. Ogale, T. Venkatesan, R. L. Greene

Center for Superconductivity Research, Department of Physics, University of Maryland, College Park, MD 20742.

A.J.Millis

Department of Physics and Astronomy, Rutgers University, 136 Frelinghuysen Rd, New Brunswick, NJ 08854.

K. Ramesha

Solid State and Structural Chemistry Unit, Indian Institute of Science, Bangalore, India

B. Hannoyer

Laboratoire d'Analyse Spectroscopique et de Traitement de Surface des Matériaux, Université de Rouen, 76821 Mont-Saint-Aignan Cedex, France

G. Marest

Institut de Physique Nucléaire de Lyon, IN2P3 et Université Lyon I, 43 Bld du 11 Novembre 1918, 69622 Villeurbanne Cedex, France

Abstract

We have investigated the structure and electronic properties of ferrimagnetic double perovskites, $A_2FeReO_6$ (A= Ca, Sr, Ba). The A=Ba phase is cubic (Fm3m) and metallic, while the A=Ca phase is monoclinic ($P2_1/n$) and nonmetallic. $^{57}$Fe Mossbauer spectroscopy shows that iron is present mainly in the high-spin (S=5/2) $Fe^{3+}$ state in the Ca compound, while it occurs in an intermediate state between high-spin $Fe^{2+}$ and $Fe^{3+}$ in the Ba compound. It is argued that a direct Re $t_{2g}$ – Re $t_{2g}$ interaction is the main cause for the metallic character of the Ba compound; the high covalency of Ca-O bonds and the monoclinic distortion (which lifts the degeneracy of $t_{2g}$ states) seem to disrupt the Re-Re interaction in the case of the Ca compound, making it non-metallic for the same electron count.

The compounds $A_2FeMoO_6$ and $A_2FeReO_6$ (A=Ca, Sr, Ba) have attracted recent attention because they may be half-metals with high magnetic transition temperatures and have spin-dependent transport properties which may be useful in magnetic devices [1-3]. These compounds were discovered in the 1960s [4-6] and are members of the broad class of $A_2BB'O_6$ 'double perovskites' [7]. Their conductivity properties are remarkable: $Ba_2FeReO_6$ compound has been known since the pioneering studies of Sleight and Weiher to have a ferrimagnetic metallic ground state [8], but our recent investigation [3] has shown that the $Ca_2FeReO_6$ compound, which has a smaller magnetization and should have the same conduction band filling, is an insulator. Although Kobayashi and coworkers [1,2] who investigated the strontium members, $Sr_2FeMoO_6$ and $Sr_2FeReO_6$, seem to regard that half-metallic ground state is a generic feature of double perovskites, evidence contradicting this viewpoint exists. More generally, a survey of double perovskite $A_2BB'O_6$ compounds shows that only very few of them exhibit both metallicity and magnetism, although several of them are ferrimagnetic. For example, among the series $Ba_2MReO_6$ (M=Mn,Fe,Co,Ni) only the M=Fe compound is both metallic and ferrimagnetic, while the M=Mn and Ni compounds are ferrimagnetic semiconductors (M=Co oxide is an antiferromagnetic semiconductor) [8]. Similarly $A_2CrMoO_6$ (A=Ca, Sr)[4], $A_2CrReO_6$ (A=Ca,Sr) [9] and $ALaMnReO_6$ (A=Ca, Sr, Ba) [10] are all ferrimagnetic but not metallic.

Changing the A-site cation from Ba to Sr to Ca drives a T=0 metal-insulator transition in the $A_2FeReO_6$ system, even though the conduction band filling is not changed. This and the widespread occurrence of insulating behavior in double perovskite systems suggest that the technologically interesting $Sr_2FeReO_6$ and $Sr_2FeMoO_6$ systems are, in some sense, 'near' a nontrivial metal-insulator transition, which must be understood before these materials can be used with confidence in applications. To shed light on this subject we have used $^{57}$Fe Mossbauer spectroscopy along with resistivity and structural determinations of the series $A_2FeReO_6$ (A=Ca,Sr,Ba). We find that the changes in the transport and structure are correlated with the changes in the Fe valence, and present an interpretation of the data.

Polycrystalline samples of $A_2FeReO_6$ (A=Ca,Sr,Ba) were synthesized as reported earlier [3] by a two-step process. First a precursor oxide of composition $A_2ReO_{5.5}$ was prepared by reacting stoichiometric mixtures of $ACO_3$ and $Re_2O_7$ which were then reacted with required quantities of Fe and $Fe_2O_3$ in sealed evacuated silica tubes at 910 C for several days until single phase products were obtained. In the final stages, pellets were sintered in sealed silica tubes at 950 C. Powder XRD data were collected using a Siemens D5005 diffractometer and Rietveld refinements were carried out using the FULLPROF program (9). $^{57}$Fe Mossbauer spectra were recorded at room temperature and at 77 K using transmission Mossbauer set up, and the resistivity was measured by the standard four-probe method.

Powder x-ray diffraction (XRD) shows that all the three $A_2FeReO_6$ (A=Ca, Sr, Ba) oxides crystallize in ordered double perovskite structures, as reported previously [8], the A=Ba and Sr phases adopting cubic (Fm3m) symmetry and the A=Ca phase, monoclinic ($P2_1/n$) symmetry. We have refined the structures of $Ba_2FeReO_6$ and $Ca_2FeReO_6$ from powder XRD data (Fig. 1 and 2).

The lattice parameters and the relevant bond lengths and bond angles are listed in Table-I. Refinements of occupancy at B and B' sites indicate that the ordering of Fe and Re atoms is 98% and 95% respectively for the Ba and Ca oxides, showing that the B-site disorder is not important for the present discussion. Larger amounts of B-site disorder have been shown to influence the properties [12]. The monoclinic distortion in the Ca-compound is a clear manifestation of the smaller size and smaller tolerance factor t (0.963) for $Ca_2FeReO_6$ than for $Ba_2FeReO_6$ (t=1.06). The monoclinic structure is the preferred structure for rocksalt type ordering of double perovskites with smaller t values, while the cubic (Fm3m) structure is preferred by double perovskites containing Ba (t~1.0) [7]. A major consequence of the distortion is that the Fe-O-Re angle which is linear (180°) in the Ba compound is considerably bent to ~ 156° in the Ca compound. Also, the octahedral Fe-O and Re-O bond lengths which are equal in the Ba-compound, become unequal in the Ca compound (Table-I). The volume per $A_2FeReO_6$ formula unit decreases from 65.25 Å$^3$ to 57.34 Å$^3$. Electrical resistivity measurements reveal that the Ba and Sr oxides are metallic and the Ca oxide is nonmetallic [3].

The powder XRD pattern of $Sr_2FeReO_6$ shows a single-phase double perovskite with a = 7.862 A°. Since both the size (1.44 A°) and A-O covalency effect of $Sr^{2+}$ are intermediate between those of $Ca^{2+}$ and $Ba^{2+}$, it is not surprising that the properties of the Sr compound are also intermediate showing metallic or semiconducting behavior depending on the sample history [2].

In order to understand the difference in the behavior more clearly we have performed $^{57}$Fe Mossbauer spectroscopy. At 77K, both $Ba_2FeReO_6$ and $Ca_2FeReO_6$ show the characteristic six-finger pattern (Fig. 3) of a ferro(i)magnet, but with different isomer shift $\delta$ (with respect to $\alpha$-Fe) and hyperfine field $H_I$: For the Ba compound, $\delta$ and $H_i$ are 0.86 mm/s and 43.7 T respectively, and the corresponding values for the Ca compound are 0.67 mm/s and 53.4 T. The values for the Sr compound are 0.71 mm/s and 47 T, which are intermediate between those for the Ba and Ca compounds. The data for $Ba_2FeReO_6$ are comparable to those previously reported [8]. Mossbauer data for $Ca_2FeReO_6$ and $Sr_2FeReO_6$ have not been reported earlier. The Mossbauer data for $Ba_2FeReO_6$ have been interpreted as indicating an intermediate valence state between high spin $Fe^{3+}$ and $Fe^{2+}$, consistent with the metallic nature of this material [8]. The distinctly smaller isomer shift of $Ca_2FeReO_6$ clearly indicates a more positive oxidation state for iron [13] than in $Ba_2FeReO_6$. Further, the hyperfine field of the Ca compound is ~20% more than the value for the Ba compound, which suggests that the electron spin of Fe is larger in the Ca compound than the Ba case. Taken at face value, the measurement implies that the spins of the two compounds are approximately in the 5:4 ratio, which means that iron is almost exclusively in the high-spin (S=5/2) $Fe^{3+}$ state in $Ca_2FeReO_6$, while in $Ba_2FeReO_6$ it is more close to high-spin $Fe^{2+}$ (S=2). Mossbauer parameters of the Sr-compound are intermediate between the two, although closer to the latter in terms of the valence state.

We now turn to the interpretation of the data. Band theory calculations [1,2] suggest that the ground state of the Sr compound is ferrimagnetic metal. The majority spin band structure is insulating (gap at the chemical potential), the Fe 3d states are fully occupied, and the Re 5d states are empty, leading to a S=5/2 ferromagnetic moment on each Fe site. The minority spin band is metallic (no gap at the chemical potential) and the occupied part of the band is composed mainly of Re d-states of $t_{2g}$ symmetry. In simple terms, one has a set of S=5/2 Fe states which polarize the Re band via what one may think of as a superexchange interaction. The published band structure calculations yield a reasonable bandwidth (~0.5 eV) for the minority spin Re band [2], but the available papers do not say whether the bandwidth arises from a direct Re-Fe-Re hopping or from a direct Re-Re overlap. Our attempts to fit the published band structure to simple tight binding models suggest that the Re-Re hopping is the dominant effect [14]. Such a hopping is not unreasonable, given that we are dealing with $t_{2g}$ orbitals which point from one Re to another (along the face diagonal of the cubic perovskite structure) and the 5d orbitals which are spatially extended [15].

With this in mind, we turn to the chemical effects of different A-site cations. $Ca^{2+}$ is distinguished from $Ba^{2+}$ not only by its smaller size (1.34 Å vs 1.61 Å for twelvefold coordination), but also by the greater covalent nature of the Ca-O bond, since it is known (11) that the acidity of alkaline earth cations increases in the order Ba<Sr<Ca. The smaller size of $Ca^{2+}$ distorts the structure of $Ca_2FeReO_6$ to monoclinic. In the distorted structure, the Fe-O-Re angle is bent from $180^o$ to ~$156^o$. This bending decreases the overlap of $e_g$ Fe and Re levels with $O_{2p\sigma}$, but might not affect the mobile $t_{2g}$ electrons so much. In the materials of interest the Re $e_g$ levels are empty, so it is difficult to see how this influences the physics. However, it would reduce the Re-Re overlap, by misaligning the $t_{2g}$ orbitals. Moreover, the greater covalency of Ca-O bonds would directly compete with the $t_{2g}$ orbitals of the transition metal B, B' cations for the O: $2p_\pi$ electrons and also reduce the Re-Re overlap. However, it is not clear that this would reduce the Fe-Fe superexchange, since the change in symmetry might offset the loss of electrons.

A further important issue is the monoclinic distortion induced by Ca. This will lift the degeneracy of the $t_{2g}$ levels on the Re site, making an insulating state easier to form in the Re sublattice. Indeed, because there are two $t_{2g}$ electrons per Re, a large enough splitting would lead to an insulating state.

The magnetoresistance of these systems shows an interesting behavior. The Ca compound did not show any significant MR even at high fields. Hence, in Fig. 4, we compare only the MR at different temperature for the Sr and Ba compounds. It is clear that the Sr compound shows a significantly higher MR than the Ba compound under comparable conditions. Kobayashi et al [2] attribute the MR of $Sr_2FeMoO_6$ to intergrain tunneling process based on an indirect inference derived from the effects of annealing on the MR behavior. If grain boundary is indeed the origin on large MR, it would be interesting to explore as to why the transport across the grain boundary is less tunable in the Ba compound as compared to the Sr compound. Another possibility is that the higher MR or a higher magnetic tunability of the kinetic energy of electrons in the Sr compound could be primarily due to its being on the metal insulator boundary between the metallic Ba compound and the insulating Ca compound. Further work is needed to sort out these issues.

In conclusion, the insulating and metallic behavior of $Ca_2FeReO_6$ and $Ba_2FeReO_6$ respectively appears consistent with the Mössbauer data which shows almost exclusively a high spin $Fe^{3+}$ state for the Ca compound and a mixed high spin $Fe^{2+}/Fe^{3+}$ state for the Ba compound. However, it is paradoxical because the Ca compound has a higher $T_C$ – one associates a higher $T_C$ with a stronger exchange interaction which comes from a greater virtual hopping and thus more covalency. We have argued that the crucial issues are actually the Re-Re overlap and the lattice distortion which lifts the degeneracy of the Re $t_{2g}$ levels, and this seems to give the insulating character to $Ca_2FeReO_6$ at two 5d electrons per Re. It must be mentioned that the occurrence of metallic and insulating states for isoelectronic series of $A_2BB'O_6$ double perovskite as a function of A site cation has not been reported to our knowledge, although it is known among isoelectronic series of $ABO_3$ oxides, for example $RNiO_3$ (R=rare earth) [17].

Acknowledgement : This work was supported by NSF-MRSEC under grant number DMR-96-32512 and DMR 9705482 (AJM). The work at Bangalore was supported by the Department of Science and Technology, Govt. of India.

References :

* On leave from Solid State and Structural Chemistry Unit, Indian Institute of Sciences, Bangalore, India

11. XRD data were collected every $0.02°$ for $10° < 2\theta < 110°$ on a Siemens D5005 diffractometer ($CuK_\alpha$ radiation). The refinements of both the structures were performed using the FULLPROF Rietveld analysis program (Rodriguez-Carvajal J., FULLPROF version 3.1, January 1996; ILL, France) and led to $R_p$ = 5.08 %, $R_{WP}$ = 6.52 %, $R_{exp}$ = 4.31 %, $\chi^2$ = 2.29, $R_{Bragg}$ = 3.67 %, $R_F$ = 6.28 % for Ba$_2$FeReO$_6$, and R$_p$ = 3.36 %, R$_{WP}$ = 4.40 %, R$_{exp}$ = 7.85 %, $\chi^2$ = 0.315, R$_{Bragg}$ = 6.2 %, R$_F$ = 7.91 % for Ca$_2$FeReO$_6$. Observed, calculated and difference profiles are given in Fig. 1.

Figure Captions :

Fig. 1 : Rietveld refinement of powder XRD data for (a) $Ca_2FeReO_6$ and (b) $Ba_2FeReO_6$.

Fig. 2 : Structures of (a) $Ca_2FeReO_6$ (left) and (b) $Ba_2FeReO_6$ (right) .

Fig. 3 : Mossbauer spectra of $Ca_2FeReO_6$ (top) and $Ba_2FeReO_6$ (bottom) at 77 K.

Fig. 4 : Comparison of the magnetoresistance behavior of (a)$Sr_2FeReO_6$ (SFRO) and (b)$Ba_2FeReO_6$ (BFRO).

Table I

Structural parameters for $Ba_2FeReO_6$ and $Ca_2FeReO_6$

( For $Sr_2FeReO_6$, a = 7.862(1) Å)

|  | $Ba_2FeReO_6$ | $Ca_2FeReO_6$ |
|---|---|---|
| Space Group | Fm3m | $P2_1/n$ |
| a( Å) | 8.0518(2) | 5.4019(2) |
| b(Å) |  | 5.5246(2) |
| c(Å) |  | 7.6847(2) |
| $β_o$ |  | 90.02(1) |
| $d_{Fe-O}$ (Å) | 2.030(3) x 6 | 1.992(3) x 2 |
|  |  | 2.033(3) x 2 |
|  |  | 1.925(3) x 2 |
| $d_{Re-O}$ (Å) | 1.996(2) x 6 | 1.971(3) x 2 |
|  |  | 1.916(2) x 2 |
|  |  | 1.986(2) x 2 |
| < Fe-O-Re | 180 | 154.2(1) |
|  |  | 156.1(2) |
|  |  | 158.6(2) |

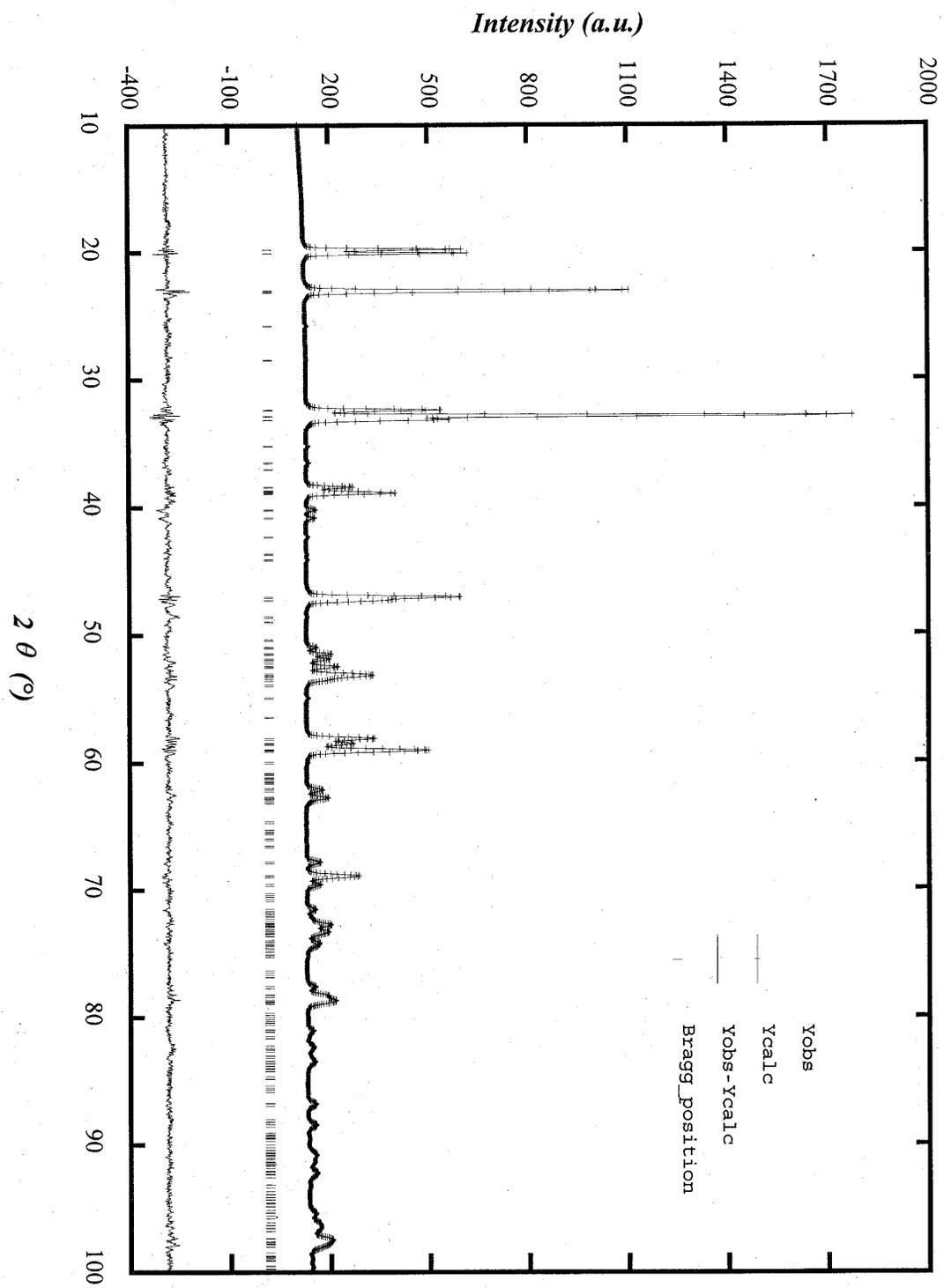

Fig 1a

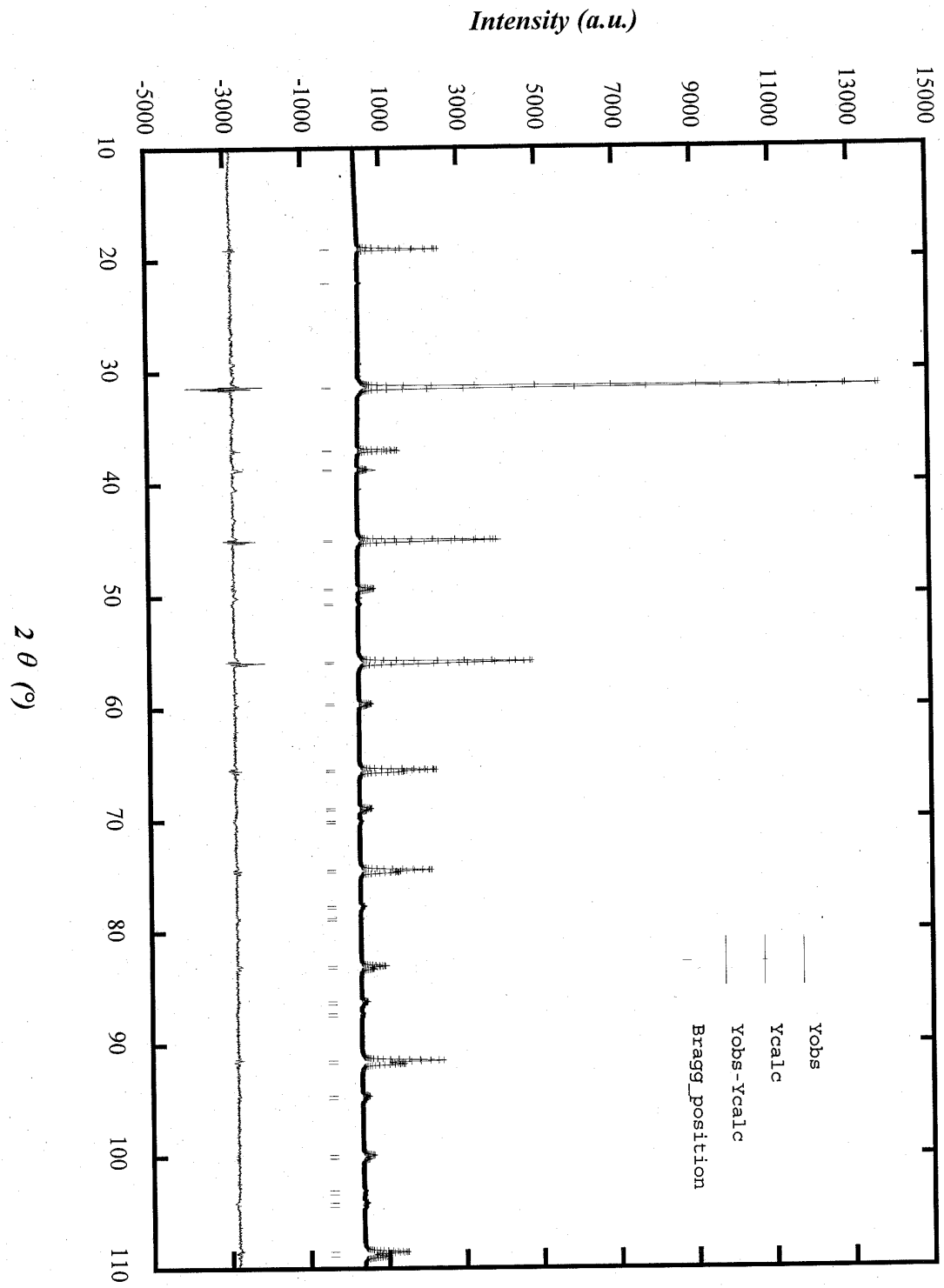

Fig 1b

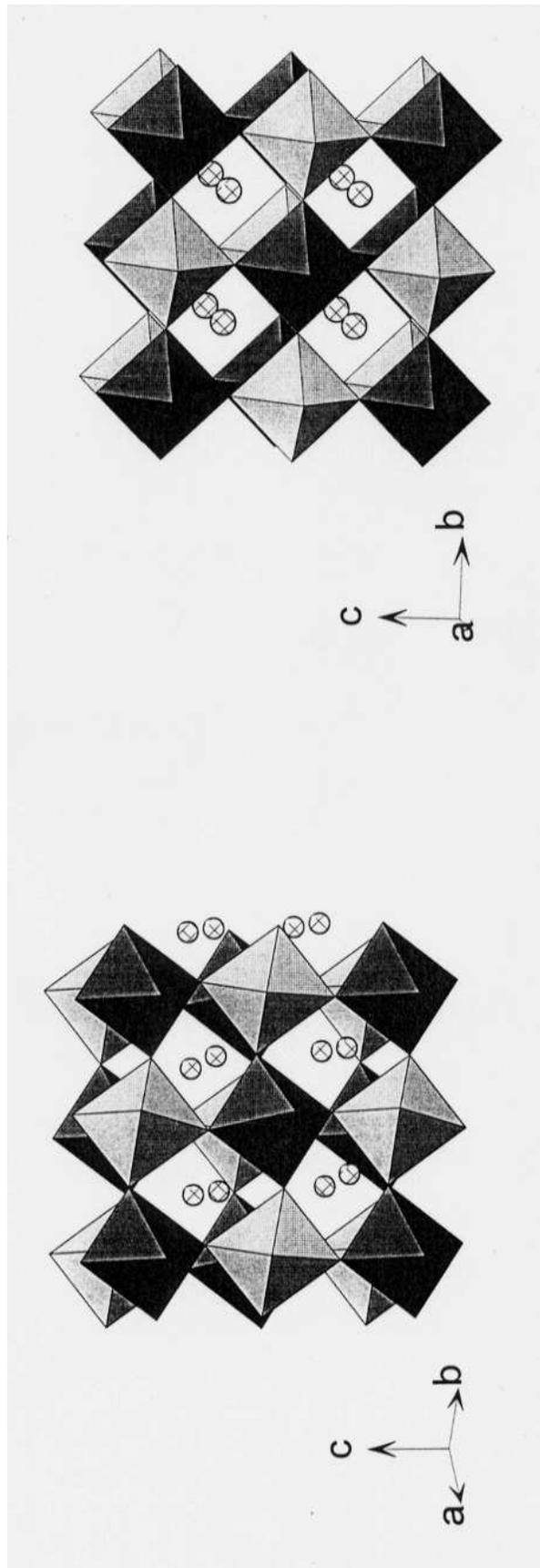

Fig 2

Fig 3

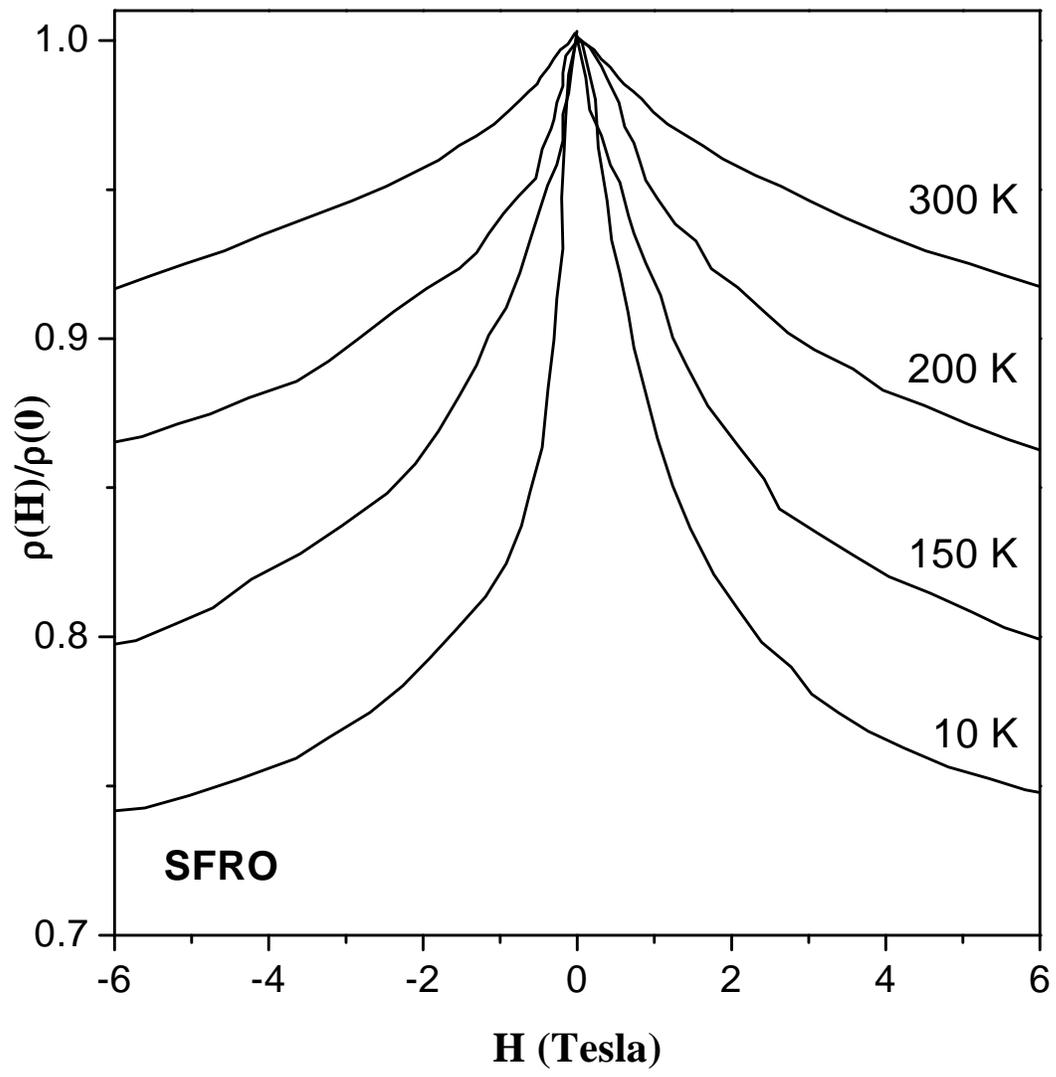

Fig 4a

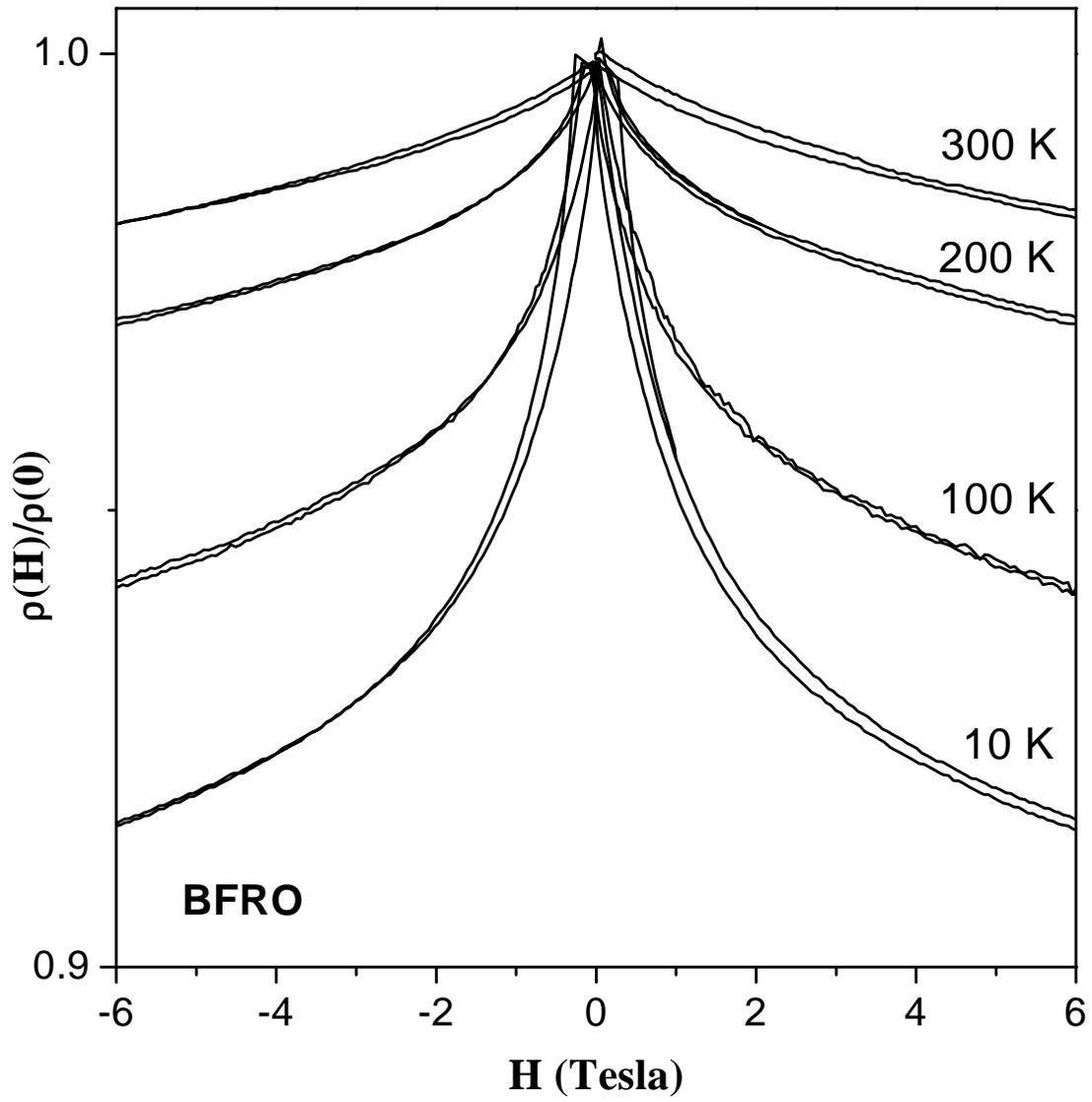

Fig 4b